# Title:

"Deep learning brain conductivity mapping using a patch-based 3D U-net"

## Authors:

Nils Hampe (corresponding author) [1,2], Ulrich Katscher [1], Cornelis A. T. van den Berg [3,4], Khin Khin Tha [5,6], Stefano Mandija [3,4]

1. Philips Research Hamburg
Roentgenstrasse 24
Hamburg, Hamburg, DE 22335

2. University Medical Center Utrecht, Division of Imaging & Oncology, Department of Radiotherapy
Heidelberglaan 100
Utrecht, NL 3582JX

3. University Medical Centre Utrecht, Centre of Image Sciences, Computational Imaging Group
Heidelberglaan 100
Utrecht, NL 3584CX
+31 88 75 506 66

4. University Medical Centre Utrecht, Imaging Division, Department of Radiotherapy
Heidelberglaan 100
Utrecht, NL 3584CX

5. Hokkaido University Hospital, Department of Diagnostic and Interventional Radiology
Kita14, Nishi5, Kita-Ku
Sapporo, Hokkaido, JP 060-8648

6. Hokkaido University, Global Station for Quantum Medical Science and Engineering, Global Institution for Collaborative Research and Education
5 Chome Kita 8 Jonishi, Kita Ward
Sapporo, Hokkaido, JP 060-0808

# Abstract:

**Purpose:**

To investigate deep learning electrical properties tomography (EPT) for application on different simulated and in-vivo datasets including pathologies for obtaining quantitative brain conductivity maps.

**Methods:**

3D patch-based convolutional neural networks were trained to predict conductivity maps from B1 transceive phase data. To compare the performance of DLEPT networks on different datasets, three datasets were used throughout this work, one from simulations and two from in-vivo measurements from healthy volunteers and cancer patients, respectively. At first, networks trained on simulations are tested on all datasets with different levels of homogeneous Gaussian noise introduced in training and testing. Secondly, to investigate potential robustness towards systematical differences between simulated and measured phase maps, in-vivo data with conductivity labels from conventional EPT is used for training.

**Results:**

High quality of reconstructions from networks trained on simulations with and without noise confirms the potential of deep learning for EPT. However, artifact encumbered results in this work uncover challenges in application of DLEPT to in-vivo data. Training DLEPT networks on conductivity labels from conventional EPT improves quality of results. This is argued to be caused by robustness to artifacts from image acquisition.

**Conclusions:**

Networks trained on simulations with added homogeneous Gaussian noise yield reconstruction artifacts when applied to in-vivo data. Training with realistic phase data and conductivity labels from conventional EPT allows for severely reducing these artifacts.

**Introduction**

Electric properties tomography (EPT) is an MR based technique that aims at deriving tissue electrical properties (EPs: conductivity σ and relative permittivity ε) by postprocessing the transmit MR radiofrequency (RF) field, known as B1+ field [1]. Since first mentioning in 1990s [2], EPT has proven clinical potentials in oncology [3,4], specific absorption rate assessment [5] and hyperthermia treatment planning [6].

According to the MR-EPT reconstruction model [1], EPs maps can be directly obtained by using the Helmholtz's equation. However, EPs reconstructions performed using commonly employed MR-EPT reconstructions methods are hampered by three major technical issues. First, the Laplacian operation is computed using numerical derivative schemes in the forms of kernels. This leads to errors at boundaries between different tissues [7,8]. Secondly, numerical spatial derivatives severely amplify the noise intrinsically present in MRI measurement [7,9]. Thirdly, the Helmholtz equation requires measurements of the complex transmit B1+ field. While measurements of the B1+ magnitude are possible with various techniques (actual flip angle method (AFI) [10], Bloch-Siegert [11] and dual refocusing acquisition mode (DREAM) [12]) measurements of the transmit RF phase are currently not feasible given clinically available MRI coil setup. The measured phase in MRI is a combination of the transmit (TX) and the receive (RX) phase, so called transceive phase (TRX). The challenge of disentangling TX and RX phase in media without specific symmetries has not been solved yet.

Previous efforts to address the aforementioned issues divide the reconstructions algorithm presented in literature into two families of algorithms. The first family based on direct methods exploits the insight that multiple measurements with different RF TX coils should yield equal EPs. By taking gradients of EPs into account, the boundary problem of conventional EPT is addressed. However, artifacts arising from patient motion and extensive acquisition time of multiple scans remain problematic. The second family corresponds to inverse methods [13]. Instead of directly estimating EPs backwards from the measured B1 map, these methods employ iterative minimization, where the forward calculated B1 maps are compared to the measured B1 maps. Thereby instable derivatives on noisy measured data are avoided. However, these methods require strong regularization to stabilize the solution in regions where the incident electric fields is low. In addition, these approaches require extensive numerical efforts when transitioned from 2D to 3D [14].

In parallel, the recent application of machine learning (ML) algorithms for image reconstructions has proven its advantages showing first promising results [15,16,17]. First attempts to use ML for EPs reconstructions have also been presented [18, 19].

In [19] local B1 phase patterns with the shape of 3D crosses were assigned EP values by finding the closest match in a dictionary. Reconstructions were post-processed by utilizing a median filter with an adaptive kernel shaped according to the corresponding MR magnitude images. Although a true learning procedure was not involved, this matching procedure corresponds to a so called k-nearest neighbor search (classification problem). Promising results showed feasibility of ML algorithms for conductivity reconstructions.

In [18], feasibility of deep learning for EPT (DLEPT) was demonstrated by using electromagnetic simulations as training data. Two experiments were conducted regarding (1) network architecture and (2) additional input data. The first compared performance of a standalone 2D U-net with an adversarial approach [20], utilizing the same 2D U-net as generator. While the latter yields visually more appealing results, highest reconstruction accuracy was attained with the standalone U-net. In the second experiment, networks employing magnitude images as input in addition to the B1 map are shown to yield improved quality EPT reconstructions for both simulated and in-vivo data. Of course, such an approach might lead to a correlation between EPs and MRI image contrast provided as a priori information to the network, which not always has to be true as they reflect different biophysical processes.

Building up on the work on training and testing deep learning algorithms for EPT (DLEPT) on simulations in [18], this work aims at answering the following question: how far are DLEPT networks able to generalize for phase-based conductivity reconstructions? In particular, is it possible to apply (a) networks trained in silico to in vivo data, and (b) to apply networks trained on volunteer data to patient data? For this purpose generalization to geometrically different tissue structures is studied by training networks on simulated and in-vivo acquired data including pathologic cases. By utilizing in-vivo B1 phase with substituted conductivity labels of conventional EPT for training, potential robustness towards artifacts related to image acquisition is tested.

**Theory**
**Conventional, Helmholtz-based EPT**
Conventional EPT is based on the Helmholtz equation (for details see, e.g., [1,5]). Assuming that variations of B1-magnitude are much smaller than variations of B1 phase $\phi^+$ allows for phase-based conductivity mapping

$$\sigma = \frac{\Delta \phi^+}{\mu_0 \omega} \quad (1)$$

with $\mu_0$ the magnetic permeability, $\omega$ the Larmor frequency, and $\sigma$ the electric conductivity. In this paper, we refer to the Helmholtz-based EPT reconstructions as: HHEPT.

The numerical solution of Eq. (1) according to [4], in connection with the widely applied 'transceive-phase assumption' [1,5], is used as reference in this study.

**Deep learning EPT**

Unlike conventional algorithms, DLEPT does not necessitate explicit formulation of the reconstruction formula. Instead, a surrogate model is learnt by minimizing the reconstruction error on training data. As this allows for the approximation of the analytically unsolvable full Helmholtz equation, direct reconstruction of inhomogeneous EPs becomes possible. Furthermore, utilization of TRX phase data for training inherently solves the transceive phase problem. Moreover, by introducing noise in the training data, noise robust reconstructions can be learnt.

DLEPT is based on fully-convolutional neural networks (CNNs), meaning that fully-connected layers are renounced completely in favor of convolutional layers. As the essential properties of CNNs (translational invariance and weight sharing) are expected to be suited for DLEPT, utilization of convolutional layers instead of fully connected layers increases network generalization by optimizing inductive bias. To allow for a mixture of features of differing scale, DLEPT reconstruction networks follow the U-net architecture, in which convolutional layers are interleaved with downsampling operations. This network is trained end to end to predict EP values from the B1 map. As up to now, no unsupervised alternative was found for DLEPT networks, supervised learning using labeled ground truth is the prerequisite for DLEPT.

**Methods**

**Network Architecture and Preprocessing:**

Empirical investigations led to the choice of a U-net with two downsampling steps (see Fig. 1). Inspired by the V-net [21], 3 convolutions are conducted in between each downsampling step instead of 2. Like the original U-net implementation and recent 3D approaches [22,23,21] in this implementation all convolutional kernels which are not used for down- or upsampling have edge length $L=3$. Convolutional kernels for down- or upsampling use $L=2$. The input stride $S_d$ for downsampling and the output stride $S_u$ for upsampling is given by $S_d=S_u=2$. Similar to several U-net implementations [22,23] all convolutions and transposed convolutions are followed by the Rectified Linear Unit (RELU) activation function. For increasing training stability and convergence speed, residual units were included in the network. Corresponding to the successful 3D U-net implementation in [23], batch normalization was added in between each convolution and the subsequent activation function. Data augmentation was performed by mirroring the input at each axis.

Unlike the 2D approach in [18], this work utilizes 3D convolutions, similar to the concept applied in [19]. This is inspired by the 3D nature of the underlying electromagnetic equations, demonstrated by the dependence of EPs on differentiation of the B1 map in all spatial directions in the Helmholtz equation (see Eq. 1).

As 3D CNNs lead to drastic increase in computational demands, limited GPU memory capacity does not allow to train full volumes. Therefore, we opted for a patch-based approach. By reducing information available to the network to the local environment this choice is expected to help prevent overfitting on global structural information. Training with patches is also beneficial in case of a small amount of ground truth data, as splitting ground truth data into multiple patches enlarges the number of samples. Empirical investigations led to an optimal patch size of 24x24x24.

To remove physically irrelevant phase offsets, the mean phase is subtracted from each phase patch. From an image processing standpoint subtracting the mean corresponds to manual guidance towards translation invariance of features. While this procedure increases network generalization, including the patch mean in preprocessing causes output dependency on patch position. Resulting artifacts at boundaries of reconstructed patches are mitigated by averaging overlapping patches. Padding of hidden layer convolutions, while introducing additional disturbances at the outer border, proved useful in reducing artifacts at patch boundaries by increasing the number of output voxels.

**Datasets:**

Three different data sets have been used in this study: a data set $DS_{sim}$ based on numerical brain simulations, a data set $DS_{vol}$ based on volunteer brain measurements, and a data set $DS_{pat}$ based on patient brain measurements. The data set $DS_{sim}$ consists of 15 realistic $B_1^+$ simulations from human brain models of the virtual population members Duke and Ella [24] with Gaussian noise added. For enlarging geometrical variability, Duke and Ella model were rigidly deformed yielding respectively 10 and 5 different brain models used for simulations (for more information about simulation dataset see [18]). These models were placed inside a virtual birdcage coil resonant at 128 MHz. B1+ magnitude and transceive phase (used for training) were obtained from 3D FDTD simulations performed in Sim4life (ZMT AG, Zurich, CH).

The two in-vivo datasets $DS_{pat}$ and $DS_{vol}$ were acquired using 3T MR scanners (Philips Healthcare, Netherlands) equipped with a quadrature RF head coil. After approval of the corresponding local Institutional Review Boards, 14 patients (mean age 42 +/- 17 yrs) with various brain lesions (see Tab. 1) were scanned at Hokkaido University (Japan) yielding $DS_{pat}$, and 18 healthy volunteers (mean age 44 +/- 7 yrs) were scanned at Philips Research Hamburg (Germany) yielding $DS_{vol}$. For all subjects, a bSSFP sequence (TR/TE=3.4/1.7ms, voxel size=1x1x1mm, flip angle=25°, 2 averages, scan duration 3:40 minutes) was acquired as suggested by [25]. HHEPT reconstructions including a subsequent bilateral denoising median filter with boundary restriction were performed according to [4].

**Experimental outline:**

At first, to evaluate mutual impact of geometrical differences, a subset of Duke-based models only was used for training the network $NW_D$ with and without added noise (SNR=100). A further network $NW_{DE}$ was trained using all Duke and Ella models, to increase geometrical variability for in-vivo testing. While above investigations demonstrate the potential of noise robustness up to SNR=100, empirical investigations led to an optimal noise level of SNR=200. Two networks were trained using the in-vivo datasets separately (network $NW_{pat}$ based on $DS_{pat}$, network $NW_{vol}$ based on $DS_{vol}$), and a third network $NW_{vol+pat}$ was trained on the combined $DS_{pat}$ and $DS_{vol}$. As these datasets are inherently affected by noise, no additional noise was added during training. Three-fold cross-validations were performed (i.e., repeatedly dividing a dataset into three groups, training the network on two of them, and applying the network to the third) using data from $DS_{pat}$ and $DS_{vol}$ separately and together. This yields training sets with 12 scans for training with $DS_{vol}$, 9/10 for training with $DS_{pat}$ and 21/22 for the combination. For investigating impact of geometrical variability, two-, three- and five-fold cross validations on $DS_{vol}$ are compared. Results are evaluated quantitatively by calculating the correlation of the reconstructed conductivity map with the HHEPT reference. Systematic differences in brain shape between $DS_{pat}$ and $DS_{vol}$ were evaluated quantitatively by calculating standard deviations of radii $sd_R$ (distances between brain surface voxels and center of mass) of individual scans.

Training and reconstructions were performed on a GPU (Nvidia GeForce GTX 1080 Ti, Santa Clara, USA) with reconstruction time of 1-2 minutes per scan. Throughout this work, all data used for validation was excluded from training.

**Results:**

Figure 2 shows reconstructions of original Duke and Ella (Duke0, Ella0) from $NW_D$. First and second column correspond to reconstructions from $NW_D$ trained without noise and with noise, respectively. Structural reconstruction errors (i.e. those exceeding single voxels) are only visible in the reconstruction of Ella0 from the network trained with noise. Quantitative analysis yields correlations of 0.991/0.947 (Duke0 for $NW_D$ without/with noise) and 0.960/0.800 (averaged over Ella0 and all Ella deformations for $NW_D$ without/with noise).

Reconstructions of scans from $DS_{pat}$ and $DS_{vol}$ using $NW_{DE}$ are compared to reference HHEPT reconstructions in Fig. 3.

Reconstructions using $NW_{pat}$, $NW_{vol}$, and $NW_{vol+pat}$ of a patient from $DS_{pat}$ and a volunteer from $DS_{vol}$, which were not included in the training set, are depicted in Fig. 4. Corresponding correlations from three-fold cross validations are presented in Tab. 2. Validations of $NW_{vol}$ included $DS_{pat}$ completely, as only volunteer folds are used for training, and analogously $NW_{pat}$ was validated on the full $DS_{vol}$.

Figure 5 depicts multiple slices from the $NW_{vol+pat}$-based reconstruction of a patient in the top row and HHEPT reference in the bottom row.

In Tab. 3, average correlations from three cross validations with different numbers of folds are shown, applying $NW_{vol}$ to $DS_{pat}$ and $DS_{vol}$. Validations included the full $DS_{pat}$.

Lower $sd_R$ =8.46±0.34 mm for $DS_{pat}$ than $sd_R$ =11.72±0.70 mm for $DS_{vol}$ corresponds to elevated sphericity of Asian patients' brains with respect to European volunteers' brains. For comparison, $sd_R$ =9.56±0.50 mm has been found for Duke and $sd_R$ =8.28±0.98 mm for Ella (including both, original and deformed models).

**Discussion**

This study investigates the generalization of DLEPT networks trained on simulations and in-vivo data with respect to differences in geometry and under the influence of simulated noise and measurement artifacts.

Networks trained on brain simulations show excellent results for the reconstruction of brain conductivity from simulated phase maps. Results degrade with (i) increasing systematic differences of brain geometry during training and testing and (ii) if noise is present during training. This indicates that noise hampers network generalization to geometries differing significantly from the training set and suggests that an exhaustive training set with different brain geometries is necessary to guarantee accurate DLEPT reconstructions.

When reconstructing in-vivo data from networks trained on simulations, additional artifacts appear especially for the cerebrospinal fluid (CSF). As these artifacts do not appear in reconstructions of simulated data, they are likely to be connected to artifacts related to image acquisition due to e.g. head motion and cardiac pulsation transferred to CSF, which affect the phase of the acquired bSSFP images [26]. However, other performance hampering influences e.g. related to electromagnetic models or simulations remain possible.

Reconstructions of in-vivo data $DS_{pat}$ and $DS_{vol}$ from networks trained on in-vivo data $NW_{pat}$ and $NW_{vol}$ show promising results, with less artifacts than in-vivo reconstructions from networks trained on simulations. This indicates that training on in-vivo data leads to reconstructions more robust towards acquisition related artifacts. Reconstruction results degrade significantly if applying $NW_{vol}$ to $DS_{pat}$ and vice versa. In fact, cross validations with different numbers of folds show that accuracy of applying $NW_{vol}$ to $DS_{pat}$ is already saturated when utilizing only half of the volunteer dataset for training. It can be concluded that differences exist between $DS_{pat}$ and $DS_{vol}$ for which the adopted networks do not generalize sufficiently. Still, the results are globally better across vol and pat if $NW_{vol+pat}$ is used. This indicates that an exhaustive dataset in training is necessary, in line with what was observed for the DLEPT reconstruction from simulated data.

In this work, in-vivo EPT reconstructions from networks trained on in-vivo data are less affected by artifacts than in-vivo EPT reconstructions from networks trained on simulated data. These unsolved challenges in reconstructing in-vivo data from networks trained on simulations are in line with results from [18], where high quality in-vivo results could only be obtained with additional image magnitude information. Nevertheless, simulations ensure a controlled environment where the performances of a neural network can be accurately tested since ground truth data is available. This is unfortunately not guaranteed for in-vivo reconstructions using networks trained with HHEPT, given the intrinsic inaccuracies of this technique. Even if accurate labels can be achieved for in-vivo data, e.g. by segmenting the tissue types and assigning respective literature values, covering an exhaustive geometrical variability in healthy and pathological tissue with in-vivo data remains challenging due to the lack of a large amount of available data. The results of this study suggest that the central advantage of using in-vivo training data is the automatic incorporation of acquisition related artifacts such as head motion and CSF pulsation. Therefore, future investigations will need to tackle the challenge of simulating acquisition related artifacts to increase robustness of networks trained on simulations and used for in-vivo reconstructions. Analogously to geometrical variability, systematic studies of generalization to artifacts differing significantly from the training set are likely to be of high value.

**Conclusion**

Neural networks were shown to be a promising candidate for solving the challenging task of extracting conductivity in-vivo without explicit edge information from the highly noise-encumbered B1 phase map. However, from the results of this study we can conclude that networks can only be applied to data with anatomic geometries and artifacts very close to the ones present in the training data. This highlights the importance of exhaustive training datasets with respect to brain geometries and tissue structures. Moreover, to exceed the conventional EPT method's accuracy, the training dataset must cover the wide variety of in-vivo artifacts as well.

**Acknowledgements**

The authors thank Niek R.F. Huttinga, Ettore F. Meliadó , Mariya Doneva, Thomas Amthor, Christian Findeklee, Christoph Leussler, Jan Hendrik Wuelbern, Alfred Mertins and Philipp Koch for their support to this work.

**References:**

[1]: Katscher U, van den Berg CAT. Electric properties tomography: Biochemical, physical and technical background, evaluation and clinical applications. NMR Biomed. 2017 Aug;30(8). doi: 10.1002/nbm.3729.

[2]: Haacke ME, Brown RW, Thompson MR, Venkatesan R. Magnetic Resonance Imaging: Physical Principles and Sequence Design, Second Edition. John Wiley & Sons. 2014; DOI:10.1002/9781118633953

[3]: Shin JW et al. Initial study on in vivo conductivity mapping of breast cancer using MRI. J Magn Reson Imaging. 2015;42:371-378.

[4]: Tha KK, Katscher U, Yamaguchi S, Stehning C, Terasaka S, Fujima N, Kudo K, Kazumasa K, Yamamoto T, Van Cauteren M, Shirato H, Noninvasive electrical conductivity measurement by MRI: a test of its validity and the electrical conductivity characteristics of glioma, Eur Radiol. 2018 Jan;28(1):348-355.

[5]: Liu J, Wang Y, Katscher U, He B. Electrical Properties Tomography Based on B1 Maps in MRI: Principles, Applications, and Challenges. IEEE Trans Biomed Eng. 2017;64:2515-2530.

[6]: Balidemaj E, Kok HP, Schooneveldt G, van Lier AL, Remis RF, Stalpers LJ, Westerveld H, Nederveen AJ, van den Berg CA, Crezee J. Hyperthermia treatment planning for cervical cancer patients based on electrical conductivity tissue properties acquired in vivo with EPT at 3 T MRI. Int J Hyperthermia. 2016 Aug;32(5):558-68

[7]: Mandija S, Sbrizzi A, Katscher U, Luijten PR, van den Berg CAT. Error analysis of Helmholtz-based MR-electrical properties tomography. Magn Reson Med. 2018 Jul;80(1):90-100.

[8]: Duan S, Xu C, Deng G, Wang J, Liu F, Xin SX. Quantitative analysis of the reconstruction errors of the currently popular algorithm of magnetic resonance electrical property tomography at the interfaces of adjacent tissues. NMR in biomedicine. 2016;29(6):744-50.

[9]: Lee SK, Bulumulla S, Hancu I. Theoretical investigation of random noise-limited signal-to-noise ratio in MR-based electrical proper-ties tomography. IEEE Trans Med Imaging. 2015;34:2220–2232.

[10]: Vasily L. Yarnykh. Actual flip-angle imaging in the pulsed steady state: A method for rapid threedimensional mapping of the transmitted radiofrequency field. Magn. Reson. Med., 57:192–200,2007.

[11]: Sacolick LI, Wiesinger F, Hancu I, and Vogel MW. B1 mapping by Bloch-Siegert shift. Magn. Reson. Med., 63(63):1315–1322, 2010.

[12]: Nehrke K and Börnert P. DREAM-a novel approach for robust, ultrafast, multislice B1 mapping: DREAMB1Mapping. Magn. Reson. Med., 68:1517–1526, 2012.

[13]: Balidemaj E, van den Berg CAT, Trinks J, van Lier AL, Nederveen AJ, Stalpers LJ, Crezee H, Remis RF. CSI-EPT: A Contrast Source Inversion Approach for Improved MRI-Based Electric Properties Tomography. IEEE Trans Med Imaging. 2015 Sep;34(9):1788-96

[14]: Leijsen RL, Brink WM, van den Berg CAT, Webb AG, Remis RF. 3-D Contrast Source Inversion-Electrical Properties Tomography. IEEE Trans Med Imaging. 2018;37:2080-2089.

[15]: Hammernik K, et al. (2018) Learning a variational network for reconstruction of accelerated MRI data. *Magn Reson Med* 79:3055–3071.

[16]: Hyun CM, Kim HP, Lee SM, Lee S, Seo JK (2017) Deep learning for undersampled MRI reconstruction. *arXiv* 1709.02576:1–11.

[17]: Zhu B1, Liu JZ, Cauley SF, Rosen BR, Rosen MS, Image reconstruction by domain-transform manifold learning, Nature. 2018 Mar 21;555(7697):487-492.

[18] Mandija S, Meliadò E, Huttinga N, Luijten P, van den Berg CAT, Opening a new window on MR-based Electrical Properties Tomography with deep learning, arXiv:1804.00016. 2018 Mar.

[19]: Hampe N, Herrmann M, Amthor T, Findeklee C, Doneva M, Katscher U, Dictionary-based Electric Properties Tomography, Magn Reson Med. 2019;81:342-349.

[20]: Isola P, Zhu J-Y, Zhou T, Efros AA, Image-to-Image Translation with Conditional Adversarial Networks. *arXiv:*1611.07004:1–16. Nov2016

[21]: Milletari F, Navab N, Ahmadi S. V-net: Fully convolutional neural networks for volumetric medical image segmentation. In 3D Vision (3DV), 2016 Fourth International Conference On, pages 565–571. IEEE, 2016.

[22]: Ronneberger O, Fischer P, Brox T. U-Net: Convolutional Networks for Biomedical Image Segmentation. ArXiv, pages 234–241, 2015.

[23]: Çiçek Ö, Abdulkadir A, Lienkamp S, Brox T, Ronneberger O. 3D U-Net: Learning dense volumetric segmentation from sparse annotation. In International Conference on Medical Image Computing and Computer-Assisted Intervention, pages 424–432. Springer, 2016.

[24]: Christ A, Kainz W, Hahn EG, Honegger K, Zefferer M, Neufeld E, Rascher W, Janka R, Bautz W, Chen J, Kiefer B, Schmitt P, Hollenbach HP, Shen J, Oberle M, Szczerba D, Kam A, Guag JW, Kuster N, The Virtual Family - Development of surface-based anatomical models of two adults and two children for dosimetric simulations. Phys Med Biol. 2010 Jan;55(2):23–38.

[25]: Stehning C, Voigt TR, Katscher U. Real-time conductivity mapping using balanced SSFP and phase-based reconstruction. Proceedings of the 19th Annual Meeting ISMRM, Montreal, QC, Canada, 2011; 128.

[26]: Katscher U, Stehning C, Tha KK. The impact of CSF pulsation on reconstructed brain conductivity. Proc. ISMRM 26 (2018) 0546

**Figures & Tables**

| lesion type | number of cases in current study |
|---|---|
| diffuse astrocytic and oligodendroglial tumors | 9 |
| meningioma | 2 |
| lymphoma | 1 |
| neuronal and mixed neuronal-glial tumor | 1 |
| hamartoma | 1 |

Tab. 1: List of lesions included in this study.

| | trained on | | |
|---|---|---|---|
| validation | volunteers | patients | volunteers + patients |
| volunteers | 0.718 | 0.525 | 0.730 |
| patients | 0.510 | 0.734 | 0.742 |

Tab. 2: Average correlations from three-fold cross validations of $NW_{vol}$, $NW_{pat}$ and $NW_{vol+pat}$.

| | num. folds/ avg. num. train. | | |
|---|---|---|---|
| validation | 2/ 9.0 | 3/ 12.0 | 5/ 14.4 |
| volunteers | 0.642 | 0.718 | 0.738 |
| patients | 0.522 | 0.510 | 0.534 |

Tab. 3: Average correlations from cross validations of $NW_{vol}$ with different numbers of folds.

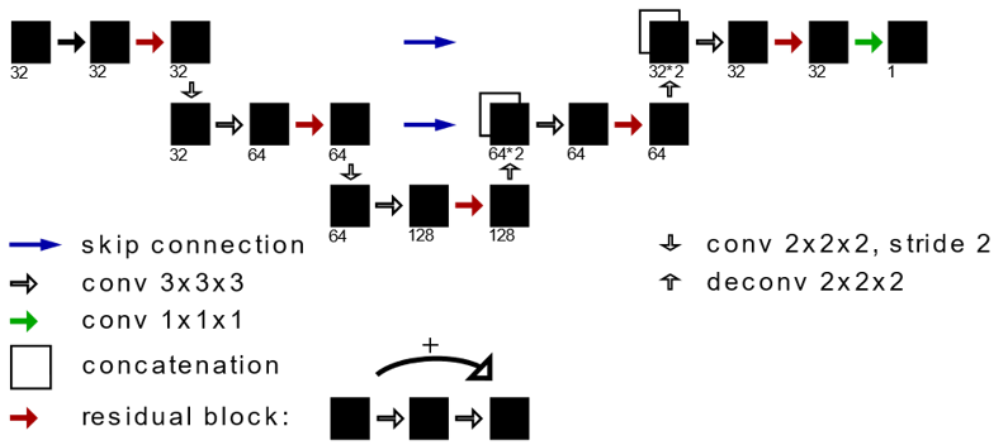

Fig. 1: Network architecture employed in this work.

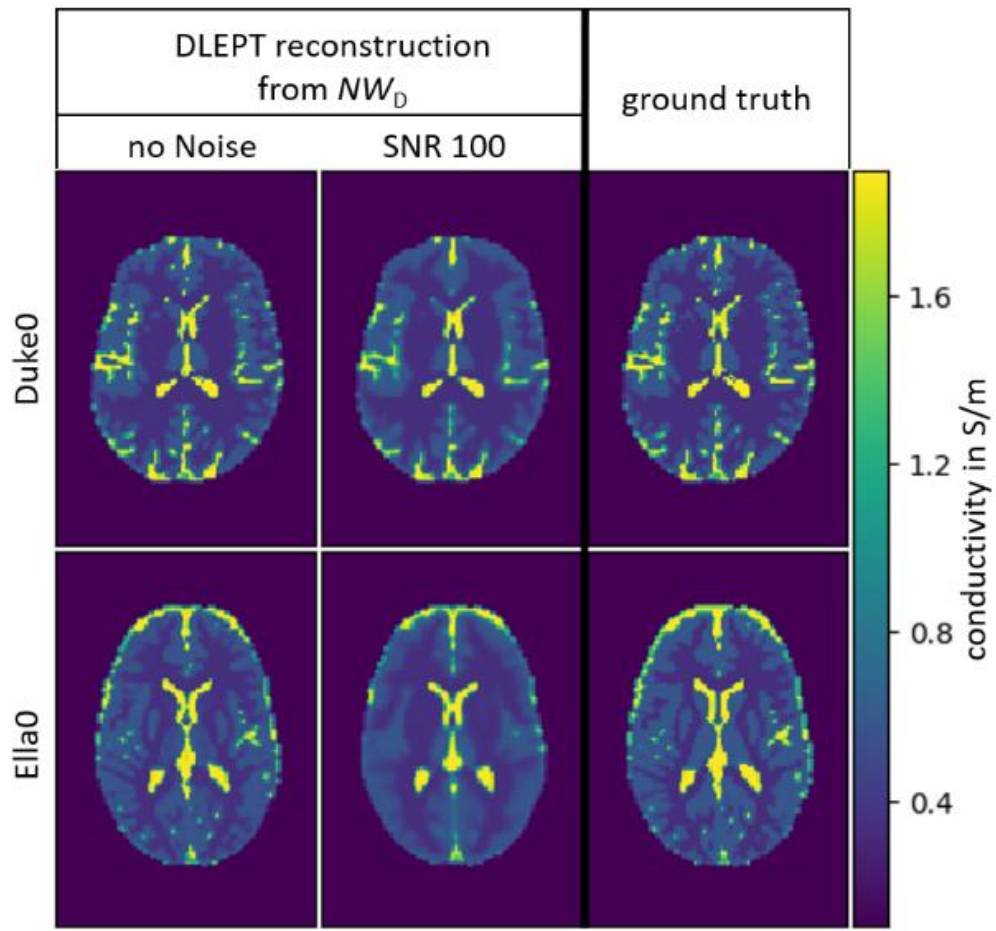

Fig. 2: Example slices of conductivity reconstructions of Duke0 and Ella0 from $NW_D$, trained with and without additive Gaussian noise.

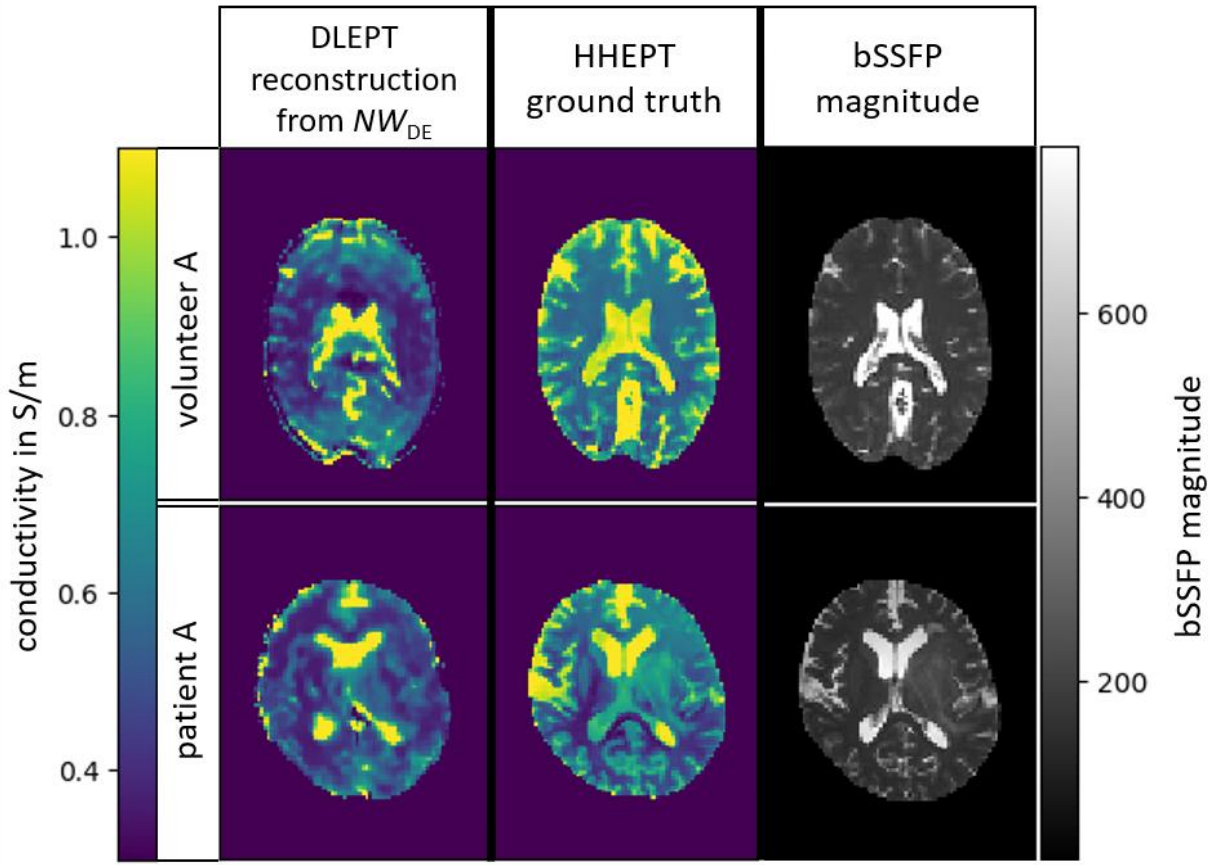

Fig. 3: Example slices of reconstructions of volunteer A and patient A by $NW_{DE}$ with HHEPT and bSSFP magnitude for reference.

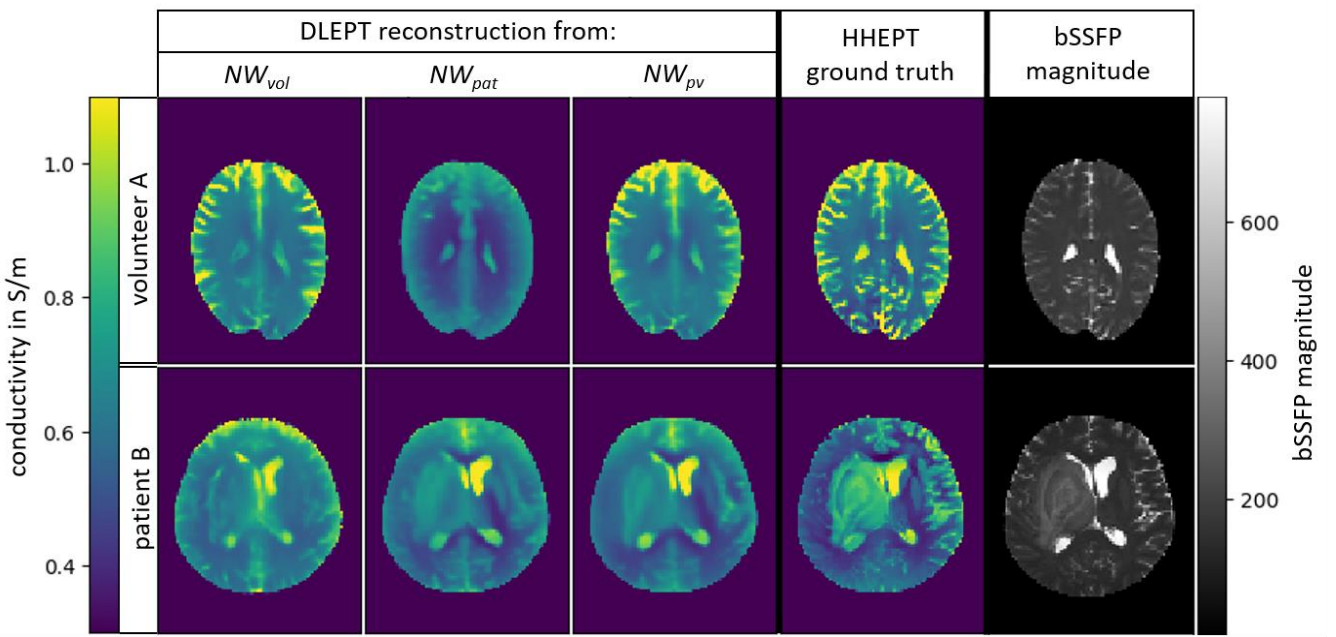

Fig. 4: Example slices of conductivity reconstructions of volunteer A and patient B from $NW_{vol}$, $NW_{pat}$ and $NW_{vol+pat}$.

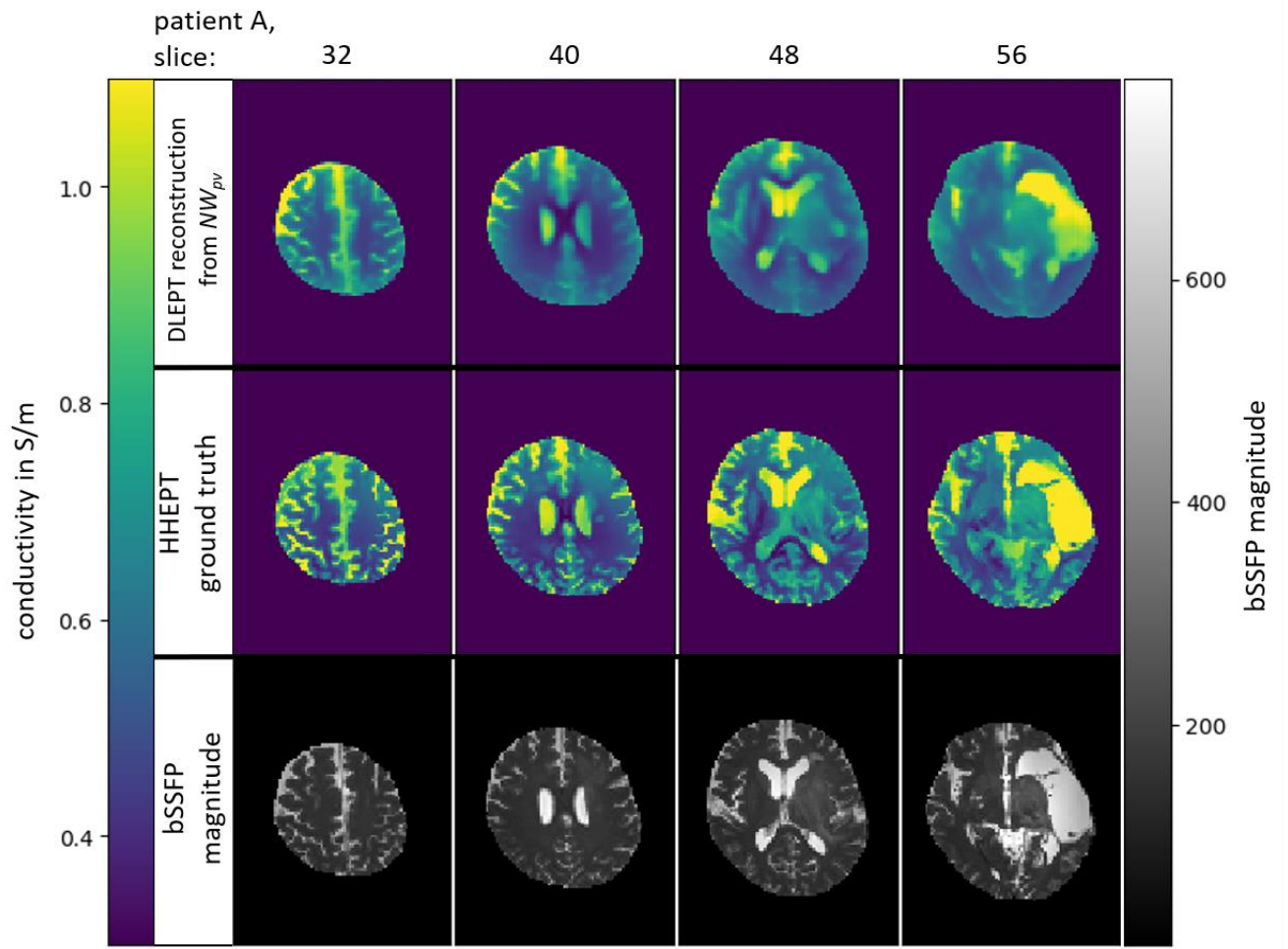

Fig. 5: Example slices of conductivity reconstruction of patient A by $NW_{vol+pat}$.